\documentclass[prb,twocolumn,showpacs,preprintnumbers,amsmath,aps]{revtex4}
\usepackage{graphicx,hyperref}
{\large {\large }}

\def\nbC{{\mathchoice {\setbox0=\hbox{$\displaystyle\rm C$}%
\hbox{\hbox to0pt{\kern0.4\wd0\vrule height0.9\ht0\hss}\box0}}
{\setbox0=\hbox{$\textstyle\rm C$}\hbox{\hbox
to0pt{\kern0.4\wd0\vrule height0.9\ht0\hss}\box0}}
{\setbox0=\hbox{$\scriptstyle\rm C$}\hbox{\hbox
to0pt{\kern0.4\wd0\vrule height0.9\ht0\hss}\box0}}
{\setbox0=\hbox{$\scriptscriptstyle\rm C$}\hbox{\hbox
to0pt{\kern0.4\wd0\vrule height0.9\ht0\hss}\box0}}}}
\def\nbQ{{\mathchoice {\setbox0=\hbox{$\displaystyle\rm
Q$}\hbox{\raise
0.15\ht0\hbox to0pt{\kern0.4\wd0\vrule height0.8\ht0\hss}\box0}}
{\setbox0=\hbox{$\textstyle\rm Q$}\hbox{\raise
0.15\ht0\hbox to0pt{\kern0.4\wd0\vrule height0.8\ht0\hss}\box0}}
{\setbox0=\hbox{$\scriptstyle\rm Q$}\hbox{\raise
0.15\ht0\hbox to0pt{\kern0.4\wd0\vrule height0.7\ht0\hss}\box0}}
{\setbox0=\hbox{$\scriptscriptstyle\rm Q$}\hbox{\raise
0.15\ht0\hbox to0pt{\kern0.4\wd0\vrule height0.7\ht0\hss}\box0}}}}
\def\nbT{{\mathchoice {\setbox0=\hbox{$\displaystyle\rm
T$}\hbox{\hbox to0pt{\kern0.3\wd0\vrule height0.9\ht0\hss}\box0}}
{\setbox0=\hbox{$\textstyle\rm T$}\hbox{\hbox
to0pt{\kern0.3\wd0\vrule height0.9\ht0\hss}\box0}}
{\setbox0=\hbox{$\scriptstyle\rm T$}\hbox{\hbox
to0pt{\kern0.3\wd0\vrule height0.9\ht0\hss}\box0}}
{\setbox0=\hbox{$\scriptscriptstyle\rm T$}\hbox{\hbox
to0pt{\kern0.3\wd0\vrule height0.9\ht0\hss}\box0}}}}
\def\nbS{{\mathchoice
{\setbox0=\hbox{$\displaystyle     \rm S$}\hbox{\raise0.5\ht0%
\hbox to0pt{\kern0.35\wd0\vrule height0.45\ht0\hss}\hbox
to0pt{\kern0.55\wd0\vrule height0.5\ht0\hss}\box0}}
{\setbox0=\hbox{$\textstyle        \rm S$}\hbox{\raise0.5\ht0%
\hbox to0pt{\kern0.35\wd0\vrule height0.45\ht0\hss}\hbox
to0pt{\kern0.55\wd0\vrule height0.5\ht0\hss}\box0}}
{\setbox0=\hbox{$\scriptstyle      \rm S$}\hbox{\raise0.5\ht0%
\hboxto0pt{\kern0.35\wd0\vrule height0.45\ht0\hss}\raise0.05\ht0%
\hbox to0pt{\kern0.5\wd0\vrule height0.45\ht0\hss}\box0}}
{\setbox0=\hbox{$\scriptscriptstyle\rm S$}\hbox{\raise0.5\ht0%
\hboxto0pt{\kern0.4\wd0\vrule height0.45\ht0\hss}\raise0.05\ht0%
\hbox to0pt{\kern0.55\wd0\vrule height0.45\ht0\hss}\box0}}}}
\def\nbZ{{\mathchoice {\hbox{$\sf\textstyle Z\kern-0.4em Z$}}
{\hbox{$\sf\textstyle Z\kern-0.4em Z$}}
{\hbox{$\sf\scriptstyle Z\kern-0.3em Z$}}
{\hbox{$\sf\scriptscriptstyle Z\kern-0.2em Z$}}}}

\begin{document}

\title{Influence of system size on the properties of a fluid adsorbed in a nanopore: Physical manifestations and  methodological consequences}

\author{Jo\"el Puibasset} \email{puibasset@cnrs-orleans.fr}
\affiliation{CNRS, Universit\'e d'Orl\'eans, FRE 3520, CRMD, 1b rue de la F\'erollerie, 
45071 Orl\'eans Cedex 02, France}

\author{Edouard Kierlik} \email{edouard.kierlik@upmc.fr}
\affiliation{LPTMC, CNRS UMR 7600, Universit\'e Pierre et Marie Curie,
bo\^ite 121, 4 place Jussieu, 75252 Paris Cedex 05, France}

\author{Gilles Tarjus} \email{tarjus@lptl.jussieu.fr}
\affiliation{LPTMC, CNRS UMR 7600, Universit\'e Pierre et Marie Curie, 
bo\^ite 121, 4 place Jussieu, 75252 Paris Cedex 05, France}

\date{\today}

\begin{abstract}
We consider the theoretical description of a fluid adsorbed in a nanopore. Hysteresis and discontinuities in the isotherms in general 
hampers the determination of equilibrium thermodynamic properties, even in computer simulations. A proposed way around this has been  
to consider both a reservoir of small size and a pore of small extent in order to restrict the fluctuations of density and approach a classical 
van der Waals loop. We assess this suggestion by thoroughly studying through density functional theory and Monte Carlo simulations the 
influence of system size on the equilibrium configurations of the adsorbed fluid and on the resulting isotherms. We stress the importance 
of pore-symmetry-breaking states that even for modest pore sizes lead to discontinuous isotherms and we discuss the physical relevance 
of these states and the methodological consequences for computing thermodynamic quantities.
\end{abstract}

\maketitle

\section{Introduction}

Phase equilibria and transitions in confined fluids have attracted wide interest over the years
among theorists and experimentalists. A ubiquitous feature of these systems is the appearance of
a reproducible hysteresis loop in the adsorption and desorption isotherms. More generally, a variety of 
intermediate states may be visited by the fluid, depending on the characteristics of the experimental 
setup. The thermodynamic analysis of these states can be quite involved due to the irreversibility which prevents  
using standard methods such as thermodynamic integration. Several
techniques have therefore been devised to calculate equilibrium properties such as the free energy, which is of primary importance 
to determine true equilibrium states and limits of metastability, or the critical nucleus for condensation in the framework of
the classical nucleation theory. Some years ago, Neimark and coworkers\cite{neimark02} proposed the ``gauge-cell method'' which
is based on the construction of a {\it continuous} isothermal trajectory of equilibrium states in the 
form of a van der Waals loop. The unstable states, which cannot be obtained in grand-canonical simulations of large systems in the 
thermodynamic limit, are then supposed to be stabilized by suppressing the density fluctuations in the system. This could be 
done by reducing both the reservoir size (which limits the fluctuations of the mean density by approaching the canonical condition) 
and the system size (which limits the local variations of density in space), which then amounts to considering a ``mesoscopic
canonical ensemble''. Provided that a continuous isotherm is obtained, equilibrium thermodynamic functions can then be determined by 
thermodynamic integration along the metastable and unstable regions.\cite{neimark02}

In our previous work,\cite{puibasset09} we studied by simulations in the mesoscopic canonical ensemble the influence of the reservoir 
size on the path followed by a simple fluid adsorbed in a nanopore of ideal geometry (a slit) during adsorption and desorption. 
We found states of intermediate density between the gas-like and liquid-like branches. The intermediate states 
break the symmetry of the slit-pore geometry in one direction, taking the morphology 
of liquid bridges (or ``rails''), gas bubbles (or gas ``cylinders'') and of undulations or bumps of the liquid layers. These states 
are ``conditionally stable'', as they are stabilized by the constraint of fixed total 
number of particles in the pore-plus-reservoir system, but we showed that they are neither stable nor metastable under 
grand-canonical conditions. (This is quite different from the situation encountered in disordered porous materials where 
the system under a strict canonical or a mesoscopic canonical constraint passes through states that are 
grand-canonically metastable.\cite{kierlik09}) We also found that discontinuities,  
which correspond to ``morphological transitions'' between different types of 
intermediate states, remain in the isotherms even for small reservoir sizes when the isotherms become reversible.

The present paper extends our previous study by investigating the effect of the system size, defined as the 
size of the simulation cell comprising the pore of slit or cylinder geometry, on the spatial variations of the density 
profile. We also analyze the robustness of the thermodynamic integration method. We consider a small (or even vanishingly small) 
reservoir, which ensures an almost canonical ensemble setting, as used in the gauge-cell method. 
We address the following theoretical and methodological questions:

(1) Are there conditions and system sizes for which the isotherms have the form of a continuous van der Waals-like loop? 

As already mentioned, suppressing the density fluctuations is the key: a van der Waals loop is typically obtained in mean-field 
approximations when density fluctuations are completely neglected. In the presence of an external potential, as in the 
present case of a fluid in a slit- or cylinder-shaped pore, this amounts to considering a mean density profile 
that retains the spatial symmetry of the pore: $\rho( z)$, with $z$ the axis perpendicular to the parallel walls in a slit pore, or 
$\rho( r)$, with $r$ the radial coordinate in the cross-section for a cylindrical pore. Calculations of this kind can be 
performed within mean-field density functional theory which has become a standard tool in the field of adsorption. 

What happens when one lifts the symmetry requirement on the density profile and allows profiles that break the pore 
symmetry? We answer this question by investigating the behavior of an adsorbed fluid both by density functional theory 
for a coarse-grained lattice-gas model and by Monte Carlo simulations of an atomic model in conditions equal or close to a 
canonical-ensemble setting. The density-functional theoretical study shows, as somewhat 
anticipated, that for a system size reduced to one lattice spacing in the two directions parallel to the slit walls (with 
periodic boundary conditions), one indeed recovers a mean density profile $\rho(z)$ and a continuous  van der Waals loop. 
However, when the system size is increased, symmetry-breaking density profiles, of the form $\rho(x,z)$ and $\rho(x,y,z)$, 
appear as equilibrium solutions for certain regions of the isotherms. These spatially nonsymmetric solutions appear for 
modest system sizes and their presence gives rise to morphological transitions and 
discontinuities in the fluid isotherms. A similar phenomenon is observed in the Monte Carlo simulation of an atomic fluid in 
slit and cylinder pores, for which discontinuities associated with morphological transitions appear in the isotherms 
for system sizes of the order of the pore width or diameter.

(2) What is the physical meaning of the intermediate, symmetry-breaking states obtained for finite system sizes? 

In the thermodynamic limit, all intermediate states should disappear from thermodynamic functions and observables as 
their free-energy is always higher than that of the main gas-like and liquid-like branches due to the cost of the additional 
interfaces involved. Nevertheless, they can still be relevant to describe effects that are subdominant to the bulk ones. 
In particular, one can take advantage of the stabilization of symmetry-breaking states and density profiles by finite 
system sizes (in the canonical ensemble) to extract the properties of the interfaces, among which the liquid-gas 
surface tension in the adsorbed fluid. We illustrate this for the density-functional calculation and the Monte Carlo simulation. 
Even one step further, some symmetry-breaking states serve as proxies for determining the characteristic nucleus for 
condensation in a pore and estimating the associated free-energy barriers.

(3) Is thermodynamic integration doomed to failure when discontinuities associated with morphological transitions are 
present in the isotherms?

We have found (see above) that discontinuities in the equilibrium isotherms appear for quite modest system sizes. In 
realistic models, smaller system sizes do lead to continuous isotherms, but the finite-size effects on the properties of the 
system are so large that one cannot hope to describe the thermodynamic limit in this way. The window of system sizes 
for which the isotherms are continuous but the thermodynamic properties are already those of the macroscopic system is 
therefore extremely small if not inexistent. This may seem as a serious drawback for thermodynamic integration. However, 
we have found that there is a significant range of system sizes that do not lead to observable finite-size effects on the 
thermodynamic properties and that still allow equilibration of the adsorbed fluid so that the isotherms are reversible (recall 
that hysteresis effects seem almost unavoidable both numerically and experimentally in large to macroscopic systems). For 
such sizes, the isotherms are discontinuous but the  discontinuities can be handled as some 
thermodynamic functions are still continuous across these discontinuities. We show that this is the case for Monte Carlo 
simulations of an adsorbed atomic fluid in a mesoscopic canonical ensemble, and this suggests that thermodynamic integration 
then remains a practical tool provided one appropriately chooses the system sizes on which to apply it.

The paper is organized as follows. In section II, we introduce the setup and the model,
and we describe the algorithms used in the DFT calculations and the simulation method. For the former, we have devised 
a way to obtain solutions for the density profiles that break the pore symmetry, in a canonical ensemble. 
In section III we discuss  the influence of the system size on the states visited by the fluid at equilibrium. A 
detailed analysis of the domain of existence of these states and of the associated symmetry-breaking density profiles is 
given for the DFT results. For the MC simulations, we focus on the validity of the 
thermodynamic-integration method in the mesoscopic canonical ensemble.  Finally, we conclude in section IV.

\section{Model and Methods}

We consider a fluid adsorbed in a pore of ideal geometry in setups corresponding to different thermodynamic ensembles:  
grand-canonical, canonical, and a mixed ensemble called the mesoscopic canonical ensemble.\cite{pana87,
smit89,smit89b} The latter is extensively used in the gauge cell method introduced by Neimark and coworkers.\cite{neimark00,
neimark05, neimark05b, neimark05c, vishny03} In this ensemble, the pore is in 
contact with a reservoir of finite (possibly small) relative size. The cross section of the pore 
is fixed, while its length may be changed in order to study its influence on the 
adsorption/desorption measurements. 

In this work, we use two complementary methods: the density-functional theory (DFT) on a coarse-grained lattice-gas model and 
Monte Carlo (MC) simulations of an atomic fluid. The former allows us to explore a large domain of system sizes and  to 
have an easy access to thermodynamic information, such as  the fluid grand potential. The later on the other hand provides a clear 
interpretation of the results in terms of molecular configurations. 

\subsection{Density Functional Theory}

As discussed in previous papers,\cite{kierlik02, detcheverry03} our DFT description is based on a coarse-grained 
lattice-gas model. We consider a fluid on a simple cubic lattice that is confined in a slit pore of fixed width $H$ in the $z$ direction,  
with periodic boundary conditions applied in the $x$ and $y$ directions parallel to the slit walls. Multiple occupancy of a 
site is forbidden and only nearest-neighbor interactions are taken into 
account. The starting point of our theoretical analysis is the following 
expression of the free-energy functional in the local mean-field approximation:
\begin{equation}
\begin{aligned}
\label{Eq01} 
F[\{ \rho_i \}]=&k_BT \sum_i [\rho_i \ln \rho_i +(1 -\rho_i) \ln (1 -\rho_i)]
\\& -w_{ff} \sum_{<ij>} \rho_i \rho_j - \sum_{i} \rho_i \phi_i,
\end{aligned}
\end{equation}
where $\rho_i $ and $\phi_i $ are, respectively, the thermally averaged fluid 
density and the (attractive) external field at site $i$; $w_{ff}$  denotes the 
fluid-fluid attractive interaction and the double summation runs over all 
distinct pairs of nearest-neighbor sites. In our case, $\phi_i=0$ everywhere except for sites 
near the pore walls where $\phi_i=w_{sf}$, the solid-fluid attractive interaction.

To get a reference, we first start with the grand canonical situation where fluid particles can 
equilibrate with an infinite reservoir that fixes the chemical potential $\mu$. 
Minimizing the grand-potential functional  $\Omega[\{ \rho_i \}]= F[\{ \rho_i 
\}]-\mu \sum_i \rho_i$ with respect to $\{ \rho_i \}$ at fixed $T$ and $\mu $ 
yields a set of  coupled equations,
\begin{equation}
\label{Eq02} 
\rho_i=\frac{1}{1+\exp [ -\beta (\mu +w_{ff}\sum_{j/i} \rho_j+\phi_i)]},
\end{equation}
where the sum $\sum_{j/i}$ runs over the $c=6$ nearest neighbors of site $i$.  By using a 
simple iterative method to solve these equations, one finds solutions that are 
only minima of the grand potential surface, \textit{i.e.} metastable states.
The adsorption isotherm is obtained by increasing the chemical potential in 
small steps $\delta \mu$. At each subsequent $\mu $, the converged solution at 
$\mu-\delta \mu$  is used to start the iterations. 

We next study the canonical ensemble, \textit{i.e.} when  the total number $N=\sum_i \rho_i$ of the fluid particles 
is fixed. The isotherm is then obtained by increasing $N$ in small steps $\delta N$. 
In this case, the system tries to minimize its Helmholtz free-energy $F$. 
This can be solved by the method of Lagrange multipliers. We 
consider the function
\begin{equation}
\label{Eq04} 
\bar{F}[\{ \rho_i \},\lambda,T]=F[\{ \rho_i \},T]+\lambda \{ N-\sum_i \rho_i\},
\end{equation}
where $\lambda $ is a Lagrange multiplier that has the meaning of a chemical 
potential coupled to the local densities. 
Minimizing $F$ with the constraint on the local densities amounts to simultaneously solving 
the coupled equations 
$\frac{\partial \bar F}{\partial \rho_i}=0$ and  $\frac{\partial \bar 
F}{\partial \lambda}=0$, or equivalently,
\begin{eqnarray}
\label{Eq05} 
& k_B T \ln [\frac{\rho_i}{1-\rho_i}]-\lambda-w_{ff}\sum_{j/i} \rho_j - \phi_i= 
0,\quad 1 \leq i \leq V \nonumber \\
& N-\sum_i \rho_i=0
\end{eqnarray}
where $V$ is the total number of sites. One has then to define an iterative scheme that specifies how 
the system goes from one converged solution to another as the total number of 
particles is slowly changed.  The details were given in a previous paper.\cite{kierlik09} 
If the algorithm converges, it does not necessarily converge to a local minimum, nor even to an 
extremum, as the constraint can stabilize an unstable state (stable or unstable refers here to the grand-canonical 
situation). As illustrated below, we indeed find physically acceptable configurations for the local densities ${\rho_i}$, 
which are nonetheless unstable in the grand-canonical ensemble (a condensation nucleus for instance).

One knows from a previous study\cite{maier01} that the solution space of the DFT in the canonical 
ensemble for an infinite slit pore can be very complex. We are not interested here in finding all solutions to the system of 
coupled non-linear equations in Eq. (\ref{Eq05}) but rather in understanding the physical states that typically appear along
 the adsorption path followed by the system when changing the system size. We have therefore performed 
 calculations by varying the lateral dimensions $L$ (length) along the $x$-axis and $\ell$ (width) along the $y$-axis 
 and applying periodic boundary conditions in both directions.

(1) As already mentioned in the introduction, the case $L=l=1$ corresponds to configurations with a uniform density 
inside a given layer, \textit{i.e.} to a density profile 
$\rho(z)$ that is invariant along $x$ and $y$ axes. The canonical
isotherm should be continuous and reversible, which allows safe thermodynamic 
integration as for the usual van der Waals loop for  homogeneous vapor-liquid phase 
transition.

(2) We then increase the length $L$ (keeping $\ell=1$) and search for solutions $\rho(x,z)$ that can break the spatial symmetry along the $x$-axis. However, 
since the DFT does not account for thermal fluctuations, the symmetry cannot be spontaneously broken with a simple iterative method. To remedy this, 
we temporarily modify the external potential by increasing (resp. decreasing), in the case of adsorption (resp. desorption), its value by a 
few percents for one site in each of the layers adjacent to the walls ($z=1$ and $z=H$). When convergence has been 
reached with the modified potential, we switch back to the original potential and restart the 
iteration process with the previously converged solution. With this 
procedure, one checks whether a symmetry-breaking solution can persist 
with a symmetric potential and therefore appears as an equilibrium solution in the canonical ensemble.

(3) Finally, we search for solutions $\rho(x,y,z)$ by varying both $L$ and $\ell$. We use the same methodology 
as before to promote symmetry breaking solutions: we temporarily modify  the external potential on a rectangular 
array of sites in the layers close to the walls, which favors the appearance of liquid bridges or vapor bubbles.
We choose a rectangle with dimensions $L/4 \times \ell/4$ because we observed that a modification on a too small array of 
sites does not allow the solution to break the symmetry. We also  systematically checked the stability of the 
solutions so obtained when reverting to the unperturbed symmetric potential.

Generically, the isotherms obtained from the above solutions display a hysteresis between adsorption and desorption. 
As extensively documented, hysteresis is ubiquitous in adsorption/desorption phenomena. It appears as a consequence of 
very long equilibration times associated with high free-energy barriers. However, the hysteresis found in the DFT description of 
adsorbed fluids is rather an artefact of the mean-field approximation that suppresses the thermal fluctuations 
and does not allow the system to overcome any energy barrier, whatever its magnitude. In the present problem, this represents 
an obstacle as we are interested in small system sizes for which equilibration should not be a problem in any realistic system. 
Actually, we find in the Monte Carlo simulations that equilibration does take place for the system sizes studied. To 
get around this problem, we have taken advantage of the fact that the free energy of the solutions is easily computed from 
Eq. (\ref{Eq01}) to look for the solutions with minimal free energy. The latter should then represent the equilibrium states and 
describe a reversible isotherm.

\subsection{Monte Carlo simulations}

The objective is to build the adsorption/desorption isotherms of a simple fluid 
in a porous material in equilibrium with a finite reservoir. In this work, the 
reservoir is chosen as small as possible, so that the statistical ensemble is 
essentially canonical. The reservoir 
is used as a tool to introduce or remove fluid in the system through favorable 
insertions, in order to build the isotherms. In our systems, the ratio between the reservoir size and 
the pore size is 66.7 for the slit pore and 50 for the cylinder. Due to the large 
difference in densities between the condensed fluid and the gas phase in the reservoir, 
essentially all atoms are then in the pore. 

The system and the reservoir are both in thermal equilibrium (imposed 
temperature) and in chemical equilibrium. The chemical equilibrium is achieved 
by particle exchange, the total number of particles being constant.  In order to 
obtain the adsorption isotherm, the total number of particles is increased 
stepwise. The extra particles are introduced in the reservoir, and, after a 
while, the system equilibrates, resulting in a new chemical potential $\mu$ for 
the system, and a new average density of particles $\rho(\mu)$ in the pore. After 
complete filling of the system, the desorption isotherm is obtained by 
decreasing in a similar way the total number of particles. Using various amounts of particles for 
addition or removal of the fluid in the system produces identical isotherms, 
with more or less points along it. A compromise is thus chosen between precision 
and computing time. In some cases to be described later, the  
adsorption/desorption isotherms may exhibit discontinuities or gaps. In such 
situations, we have decreased the particle increment around the discontinuities to 
improve the accuracy.

The numerical resolution is performed by Monte Carlo simulation using the 
Metropolis algorithm for particle displacements (thermalization) and particle 
exchange between the porous material and the reservoir (chemical equilibration). 
The pore geometry (slit or cylinder) and the (monoatomic) fluid are chosen to 
capture the main features of fluid states in pores of simple geometry. The 
system mimics argon adsorption in nanoporous solid carbon dioxide. The 
interactions are modeled by Lennard-Jones potentials, which are truncated at three atomic 
diameters for convenience. The parameters describing the fluid-fluid interactions 
are those for argon ($\epsilon_{\rm ff} / k_B = 119.8$ K and $\sigma_{\rm ff} = 
0.3405$ nm), while the fluid-wall parameters are $\epsilon_{\rm fw} / k_B = 153.0$ K 
and $\sigma_{\rm fw} = 0.3725$ nm.\cite{puibasset05, puibasset05b} All 
thermodynamic quantities are normalized by the fluid-fluid Lennard-Jones 
interaction parameters. The wall roughness of the pores is neglected, and the 
fluid-wall potential is obtained by integrating the Lennard-Jones potential over 
a uniform density of interacting sites.

Two pore geometries are considered. A slit pore of given height $H = 6$ (in 
units of the fluid atomic diameter) and dimensions parallel to the walls equal to $\ell = 
8$ in one direction (width) and varying between $L = 4$ and $L = 40$ in the 
other direction (length of the pore). Periodic boundary conditions are 
applied in the two directions parallel to the walls. The cylindrical pore is of 
fixed diameter $D = 6$, the length varying between $L = 4$ and $L = 24$. 
Periodic boundary conditions are applied along the cylinder axis. It has been 
checked that the gas in the reservoir is always ideal in this study, and thus 
need not to be treated explicitly, which speeds up the calculations. 

In order to discuss notions such as stability, coexistence and nucleation in 
the mesoscopic canonical ensemble, one has to calculate thermodynamic potentials. The 
whole pore-plus-reservoir system being in the canonical ensemble, the total 
Helmholtz free energy is the quantity of interest. For a small reservoir 
(quasi-canonical situation), this total free energy is close to the Helmholtz 
free energy of the fluid confined in the pore. On the other hand, if the 
reservoir is large enough, the pore is almost in the grand-canonical ensemble, 
and one is then interested in the calculation of the grand potential. The grand 
potential is also useful to check whether the (inhomogeneous) intermediate fluid 
states observed in the mesoscopic canonical ensemble are metastable in the grand canonical 
ensemble.\cite{puibasset09} The grand-potential density may be evaluated 
either directly during the course of the simulation or by thermodynamic 
integration along the adsorption isotherm. The corresponding algorithms are presented now.

In our previous work\cite{kierlik09, puibasset09} we showed that the 
chemical potential, when defined in each part of the system (pore and 
reservoir) by using Widom's insertion method or the ideal approximation in the 
reservoir, exactly coincides with the rigorous definition derived from the canonical 
ensemble for the complete system. Similarly, a thermodynamic pressure 
(denoted $\Pi_{\rm P}$), equal to minus the 
grand-potential density, may be defined in the pore  
as the conjugate quantity to the pore dimension $L$ parallel to the walls. This quantity is 
calculated numerically during the course of the simulation by using the 
virtual-volume-variation method\cite{eppenga84, harismiadis96, 
vortler00} according to
\begin{eqnarray}
\Pi_{\rm P} = \Pi_{\rm P}^{\rm id} + \Pi_{\rm P}^{\rm ex} = k_B T \frac{ \left< N 
\right> } { \mathcal{A} L} + \frac{1}{\mathcal{A}} \left< {\frac{ \delta U 
}{\delta L}} \right> {\rm ,}
\label{Eqn_direct}
\end{eqnarray}
where $\Pi_{\rm P}^{\rm id}$ and $\Pi_{\rm P}^{\rm ex}$ are the ideal and excess 
contributions to the grand-potential density, $\mathcal{A}$ is the cross-section 
area (equal to $H \ell$ for the slit pore, and $\pi D^{2}/4$ for the cylinder), 
$U$ is the configuration energy of the fluid confined in the pore, $\delta U$ 
its variation associated with the virtual variation $\delta L$ (in this case, 
a homogeneous stretching parallel to the pore length $L$), and the brackets denote 
the statistical average over Monte Carlo configurations. Note that the virtual 
changes do not interfere with the Markov chain generation, \textit{i.e.} they 
are not Monte Carlo trials to be accepted or rejected to generate new 
configurations. 

It should be stressed that Eq.~(\ref{Eqn_direct}) holds only if the fluid is 
homogeneous in the direction where the virtual variations are performed. 
The method can thus never be applied along the radial direction for the cylindrical pore, 
or perpendicular to the walls for the slit pore. The most favorable direction is along the 
pore length $L$ (pore symmetry). However, as shown later, the fluid profiles may break the pore 
symmetry: in this case the virtual-volume-variation method fails. In the particular 
case of the slit pore, due to the symmetry between the $\ell$ and $L$ directions, the 
virtual-volume-variation method can also be performed along the direction $\ell$. This 
property can be used to calculate the grand-potential density in the situations where 
the fluid profile break the pore symmetry along $L$ while it remains homogeneous along the $\ell$ direction. 

The Helmholtz free energy $F$ of the pore-plus-reservoir system can be 
calculated by integrating the relation
\begin{eqnarray}
\left. \frac{\partial F(N, V, T)}{\partial N} \right|_{ V, T} = \mu {\rm .}
\label{Eqn_Int}
\end{eqnarray}
where $N$ is the total number of particles including the pore and the reservoir. 
Note that this integration should be restricted to the continuous portions of 
the isotherms. A portion of an isotherm is qualified as continuous if it is 
possible to reduce the interval between two successive simulation points by 
performing extra simulations with smaller increments in the total amount of 
particles. In this situation, the numerical (discrete) thermodynamic integration 
of $\mu (N)$ can give the total Helmholtz free energy variations within an uncertainty  
that can be made as small as wanted by decreasing the increments between 
successive points (with, of course, the lower limit of one atom). 

In principle, this integration cannot be performed over the whole isotherm  
when the latter exhibits discontinuities. However, when the isotherms 
are \emph{reversible} (see below), and therefore presumably at equilibrium, 
one can use thermodynamic integration even when discontinuities in $\mu (N)$ are 
present, as the total Helmholtz free energy $F$ and the total number of 
particles $N$ are the same on both sides of the discontinuities.

The grand potential in the pore is then given by
\begin{eqnarray}
\Omega_{\rm P} = F - \mu N - \Omega_{\rm R}
\label{Eqn_Omg}
\end{eqnarray}
where $\Omega_{\rm R}$ is the grand potential in the reservoir. As previously 
mentioned, the gas in the reservoir is close to ideality, and therefore 
$\Omega_{\rm R} = -k_B T \left< N_{\rm R} \right>$, where $\left< N_{\rm R} 
\right>$ is the average number of particles in the reservoir. The corresponding 
grand-potential density is $\Pi_{\rm P} = - \Omega_{\rm P}/ V_{\rm P}$, where $ 
V_{\rm P}$ is the pore volume.

\section{Results}

\subsection{DFT: influence of the system size on the equilibrium density profiles}

All the illustrative calculations here are for a pore width $H=6$, a 
reduced temperature $T^*=k_BT/w_{ff}=1.0$ and a ratio $w_{sf}/w_{ff}=3$, as already 
studied in Ref.~[\onlinecite{monson08}]. They are performed by assuming the reflection symmetry, 
$\rho(x,y,z)=\rho(x,y,H-z)$.

\subsubsection{Grand-canonical isotherms}
\begin{figure*}
\resizebox{1.4\columnwidth}{!} {\includegraphics{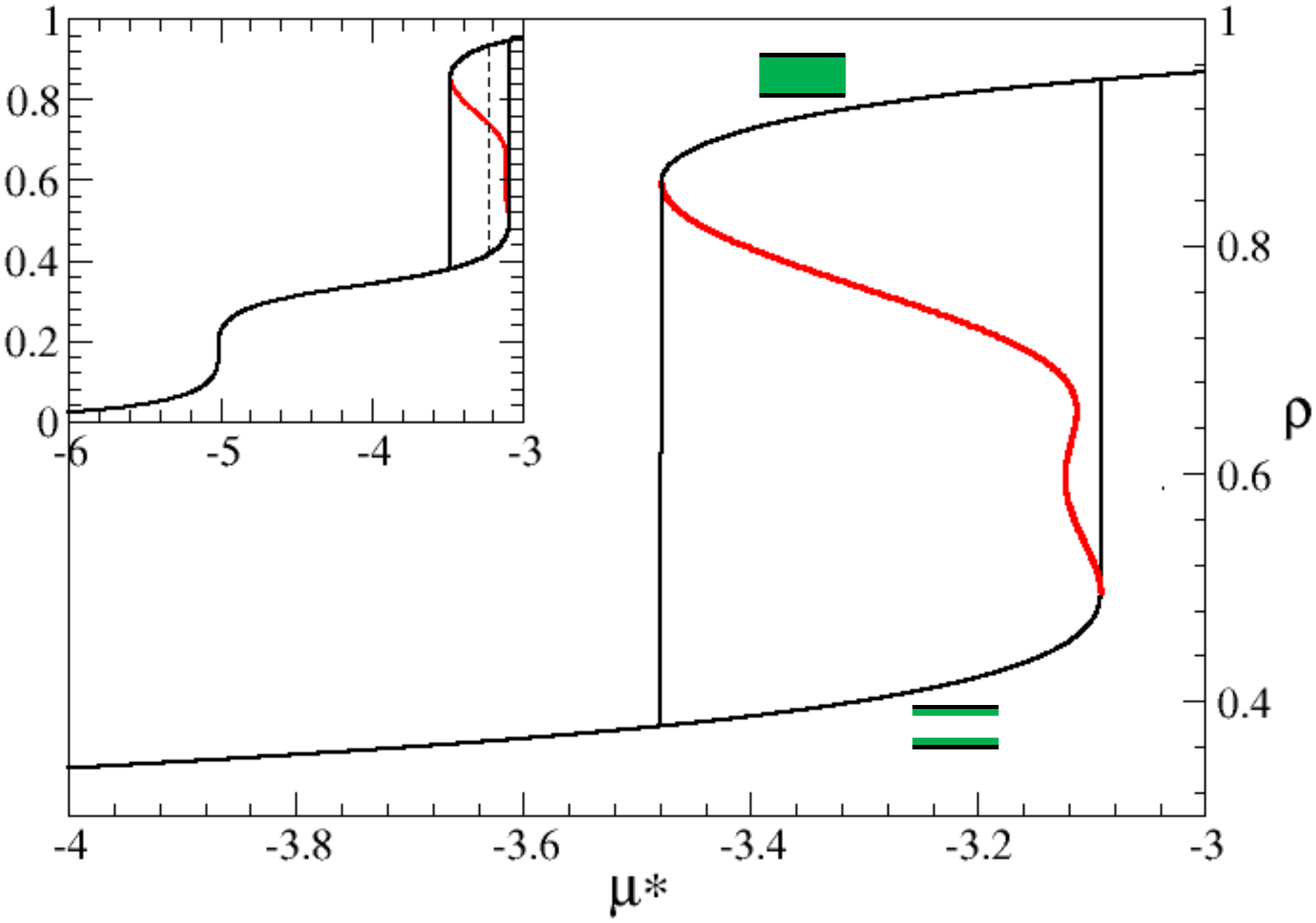}}
\caption{(color on line) DFT adsorption and desorption isotherms for a fluid adsorbed in a slit pore in the grand-canonical (black line with vertical jumps) 
and canonical (red line) ensembles. The reduced chemical potential $\mu^*$ is defined in units of $w_{ff}$. 
Inset: The whole isotherm. Main panel: The capillary-condensation region. Sketches of the fluid density distribution are also shown.}
\label{fig_LMFT1} 
\end{figure*}

The whole adsorption (desorption) grand-canonical (GC) isotherm is obtained by starting from a low 
(high) chemical potential and increasing (decreasing)  the chemical potential in 
a series of steps, with the solution at each state forming the initial guess for 
the solution at the next state. Results for the average density are shown in 
Fig~\ref{fig_LMFT1}. The isotherms exhibit two jumps corresponding, first, to a 
monolayer formation involving the sites adjacent to each of the pore walls 
and, second, to a vapor-liquid transition (capillary condensation) for the 
confined lattice fluid as illustrated in Fig.~\ref{fig_LMFT1} by sketches of 
the density distribution. This latter jump is accompanied by a pronounced 
hysteresis and the equilibrium vapor-liquid transition is determined by the 
crossing of the adsorption and desorption branches in the $(\mu, \Omega)$ plane. 
These results are what could be expected for this very simple and ideal system. 
Note that one never observes inhomogeneous density profiles in the $(x,y)$ 
plane, even when temporarily breaking the symmetry of the external potential as detailed 
in section II. More precisely, no coexistence 
between the vapor-like and the liquid-like phases is observed in the GC ensemble as such coexistence needs an 
interface and requires additional free-energy cost compared to the homogeneous vapor or 
liquid states: such states are therefore never minima of the grand potential.

\subsubsection{Evolution of the canonical isotherms varying the size of the system}
We are primarily interested in studying the solutions for the fluid in a slit pore under canonical-ensemble  [fixed 
$(N,V,T)$] conditions. In this case, the increment of the average density $\overline{\rho} = \frac{N}{V}$ has been chosen as small as $10^{-4}$ 
and convergence has been assumed when differences for $\rho$ between two successive iterations is below $10^{-8}$. 
We describe below three different situations according to the size of the system along the $x$-axis, $L$, and along the $y$-axis, $\ell$.

\paragraph{\rm$L=\ell=1$.}
One first begins with solutions of the form $\rho(z)$. As illustrated in Fig.~\ref{fig_LMFT1}, 
the sequence of states follows the GC isotherm until the end of the adsorption 
branch. Beyond the GC stability limit of the vapor-like branch, the loop exhibits 
re-entrance (decreasing $\mu$ with increasing $\overline{\rho}$) until it 
reaches the GC liquid-like branch. The (expected) continuous $S$-shape of the isotherm shows however 
a small portion with a positive isothermal compressibility, {\it i.e.} $\frac{\partial 
\overline{\rho}}{\partial \mu} > 0$, which is then grand-canonically stable. The other re-entrant 
states are grand-canonically unstable and are only stabilized by the constraint (closed system of a finite volume). 
Nevertheless, the consistency of the 
DFT guarantees that integrating $\mu(N) dN$ between $N_1$ and $N_2$ gives the 
differences $F_2-F_1$ computed from Eq.~\ref{Eq01}.

\paragraph{\rm $L>1,\ell=1$.}
When $L$ remains small, one finds the same solution as for $L=\ell=1$, {\it i.e.} the solution that preserves the symmetry 
of the external potential. Spatial fluctuations in the $x$-direction disappear when one removes the perturbative
symmetry-breaking potential. Above $L=6$ however, even if the continuous character of the isotherm is preserved, a large 
deviation appears in the upper part. Inspection of the density profiles reveals that a symmetry breaking takes place. The 
density profile is no longer uniform along the $x$-axis and shows a significant undulation outside the first layer in contact with the 
substrate. The deviation in the 
hysteresis loop from the $L=1$ case has a local $S$-shape that displays a horizontal jump for $L$ = 8 and above.

\begin{figure*}
\resizebox{1.4\columnwidth}{!} {\includegraphics{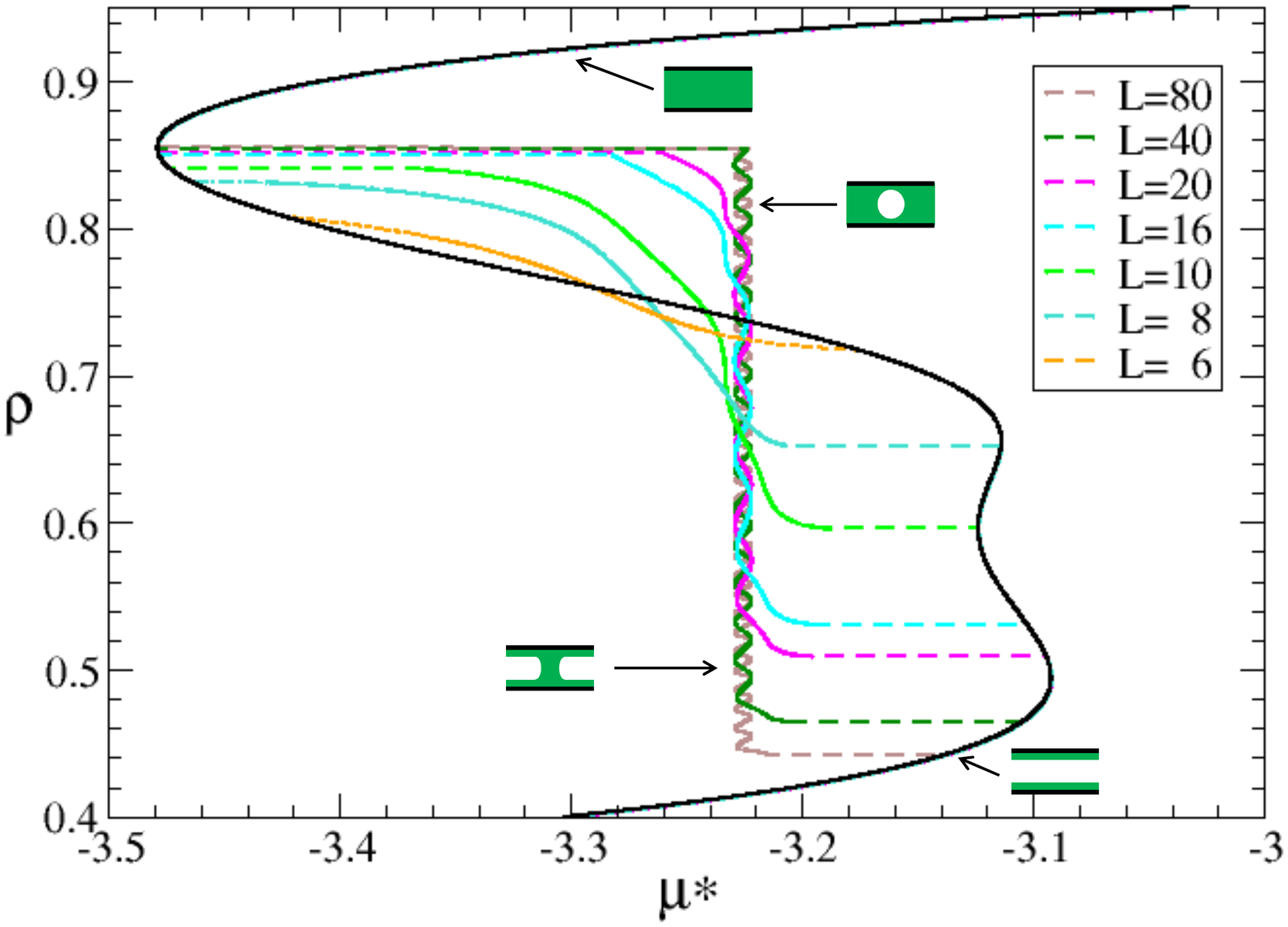}}
\caption{(color on line) DFT canonical desorption isotherms for various lengths $L$ and $\ell=1$. The density is allowed 
to fluctuate along $x$ and $z$. Sketches of the fluid density distribution are also shown in the $(x,z)$ plane.}
\label{fig_LMFT2} 
\end{figure*}

\begin{figure*}
\resizebox{1.7\columnwidth}{!} {\includegraphics{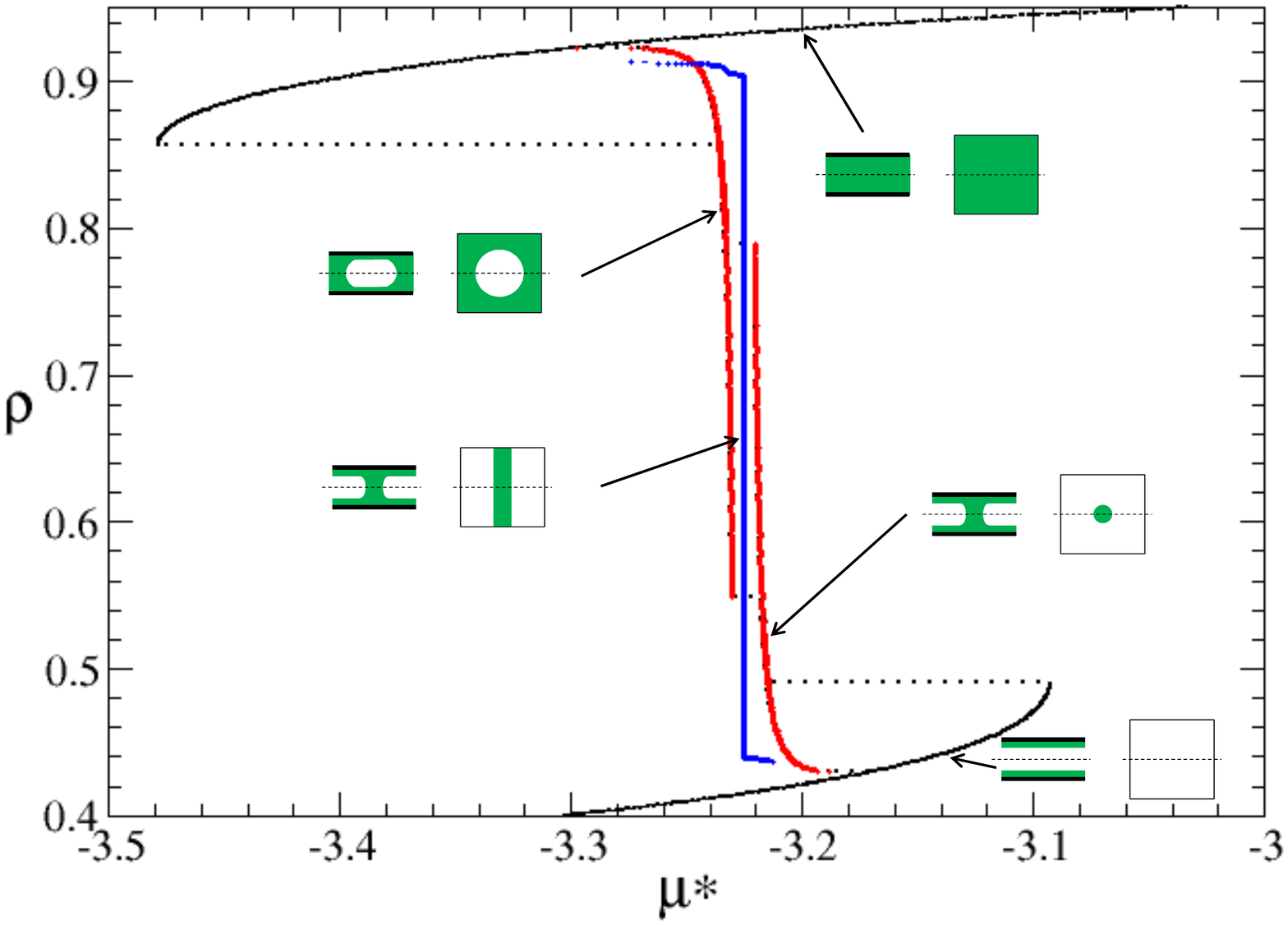}}
\caption{(color on line) Stable and metastable states for $L=\ell=100$, as obtained by DFT in the canonical ensemble. 
Two ``homogeneous'' (gas-like and liquid-like phases that satisfy the slit-pore symmetry, shown in black), and three inhomogeneous branches 
(rail, shown in blue, droplet and bubble, shown in red) are displayed with their whole domains of stability. Sketches 
of the fluid density distribution are shown both in the $(x,z)$ and $(x,y)$ planes. For clarity, the oscillations of 
the chemical potential of the rail state around the coexistence value are not displayed and only the average value is plotted.}
\label{fig_LMFT3} 
\end{figure*}


\begin{figure*}
\resizebox{1.7\columnwidth}{!} {\includegraphics{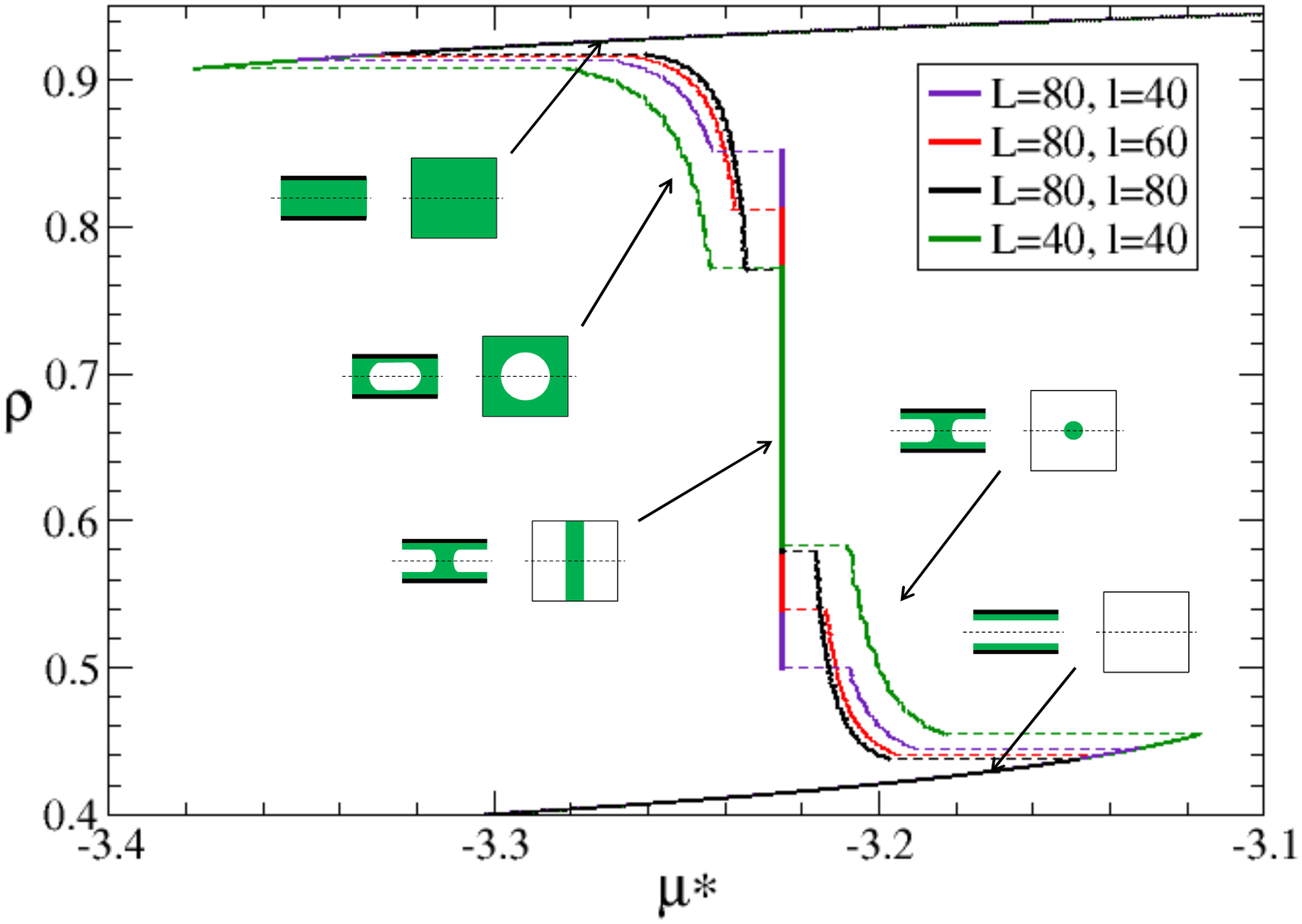}}
\caption{(color on line) Equilibrium states for various system sizes $L$ and $\ell$, as obtained by DFT in the canonical ensemble. 
Jumps between these branches are the signatures of morphological transitions. Sketches 
of the fluid density distribution are shown both in the $(x,z)$ and $(x,y)$ planes. For clarity, the oscillations of 
the chemical potential of the rail state around the coexistence value are not displayed and only the average value is plotted.}
\label{fig_LMFT5} 
\end{figure*}

This is illustrated in Fig.~\ref{fig_LMFT2} in the case of the canonical desorption 
isotherms, where the effects are more pronounced than in the adsorption case. 
Starting from saturation, all systems, whatever the size (including $L=1$), follow the GC desorption branch until the stability limit of 
the liquid-like phase on this branch. Below this point, differences appear between the solution $\rho(z)$ (corresponding to $L=1$) and the 
symmetry-breaking solutions $\rho(x,z)$. The first horizontal jump that leads to a departure from the $L=1$ canonical isotherm 
takes place for a point in the $(\mu,\rho)$ plane which becomes closer to the stability limit of the upper GC branch 
as $L$ increases. The system then reaches an inhomogeneous state which, 
above $L=40$, corresponds to a ``rail'' of vapor in a liquid environment (assuming the invariance along the $y$-axis). 
The chemical potential for this rail state corresponds to the liquid/vapor 
coexistence in the pore, $\mu_{coex} = -3.22$. (The chemical potential actually oscillates around this value, but this is an artifact 
of the underlying lattice.) This can be understood as follows. 
The rail of vapor needs a minimum size to be (canonically) stable. To generate  
this inhomogeneous state, a depletion should be created in the homogeneous (along $x$-axis) liquid 
phase and the corresponding excess density redistributed in the rest of the system. This becomes  
easier as the longitudinal system size $L$ increases and for a large $L$ it happens as soon as the 
GC branch becomes unstable. For smaller values of $L$ the presence of some extensions that stick out of the 
liquid-vapor vertical coexistence line reflects the fact that the hole has not yet reached its canonically 
stable shape and is deformed. 

The same type of behavior occurs for the second jump in the desorption isotherm, from a 
rail of liquid in a vapor environment to the lower (vapor-like) GC branch. 
As discussed in a previous paper,\cite{puibasset09} this second jump should disappear in the 
thermodynamic limit as the system would then follow the liquid-vapor coexistence line down to  
the GC branch.
For the canonical adsorption isotherms (not shown), the phenomena are reversed: for large 
systems, the first jump links the end of the lower GC branch to the liquid-vapor coexistence line and 
the second jump gets closer and closer to the upper (liquid-like) GC branch as the system size increases 
until it vanishes in the thermodynamic limit.

\paragraph{\rm$L>1,\ell>1$.}
Increasing the size of the simulation box in both lateral directions and allowing fluctuations along the $y$-axis in addition 
to those along the $x$-axis reveals the existence of other inhomogeneous metastable states.
This is illustrated in Fig.~\ref{fig_LMFT3} for the largest system 
($L=100, \ell=100$) that we have studied. The two separate branches on each side of the liquid-vapor coexistence 
line are reached by the fluid when the homogeneous states become unstable to the introduction of  
the rectangular symmetry-breaking perturbative potential. Upon adsorption (resp. desorption), 
the inhomogeneous state reached by the system corresponds to a liquid droplet 
(resp. vapor bubble) bridging the two surfaces of the slit pore in a vapor (resp. liquid) 
environment. This is easily seen by visual inspection of the density profiles. The system does not 
spontaneously reach the rail state, as discussed below, and the droplet (resp. bubble) 
branch extends until its stability limit. However, this kind of behavior is not generic: for 
smaller values of $\ell$, the adsorption (resp. desorption) isotherm jumps from the droplet 
(resp. bubble) to the rail state and a second jump to the upper (resp. lower) GC branch next takes place. 

\subsubsection{Morphological transitions and interfacial properties}

Searching for symmetry-breaking density profiles in systems of increasing size has revealed the existence of inhomogeneous states. 
Jumps take place between these states and the gas-like and liquid-like states, but not at equilibrium. Instead, a large 
hysteresis  is observed when varying the mean density, which corresponds to the ``superspinodals'' found in Ref.~[\onlinecite{neimark00}]. 
To get rid of this spurious hysteresis (see the discussion at the end of II.A), we have computed 
the equilibrium canonical isotherm for a given system size by finding the minimum of 
the Helmholtz free energy among all identified states : the two symmetric liquid-like and vapor-like (upper and lower GC 
branches) and the three inhomogeneous (rail, droplet and bubble branches) states. The jumps in 
the chemical potential between these equilibrium branches correspond to ``morphological transitions'', \textit{i.e.} a sudden 
spatial redistribution of the density profile inside the pore at a fixed mean density. Results are shown in Fig.~\ref{fig_LMFT5} 
for various system sizes ($L\times \ell = 40\times 40, 80 \times 80, 100 \times 100$, and $80 \times 40$).

\paragraph{Nucleation barriers.}
Generically, the system now visits all five states when one varies the mean density. As Everett and Haynes\cite{everett72} suggested, 
these states, which are stabilized in the canonical ensemble, could be relevant to understand the real path followed by the 
system during a vapor/liquid transition. More recently, Binder and coworkers\cite{schrader09,binder03,binder12} showed 
that the intermediate states visited by the system along the adsorption/desorption isotherm are relevant for the 
determination of the fluid interfacial properties (surface tension, Tolman length), 
which can then be used in the framework of the classical nucleation theory for 
computing nucleation barriers. 

\begin{figure*}
\resizebox{1.7\columnwidth}{!} {\includegraphics{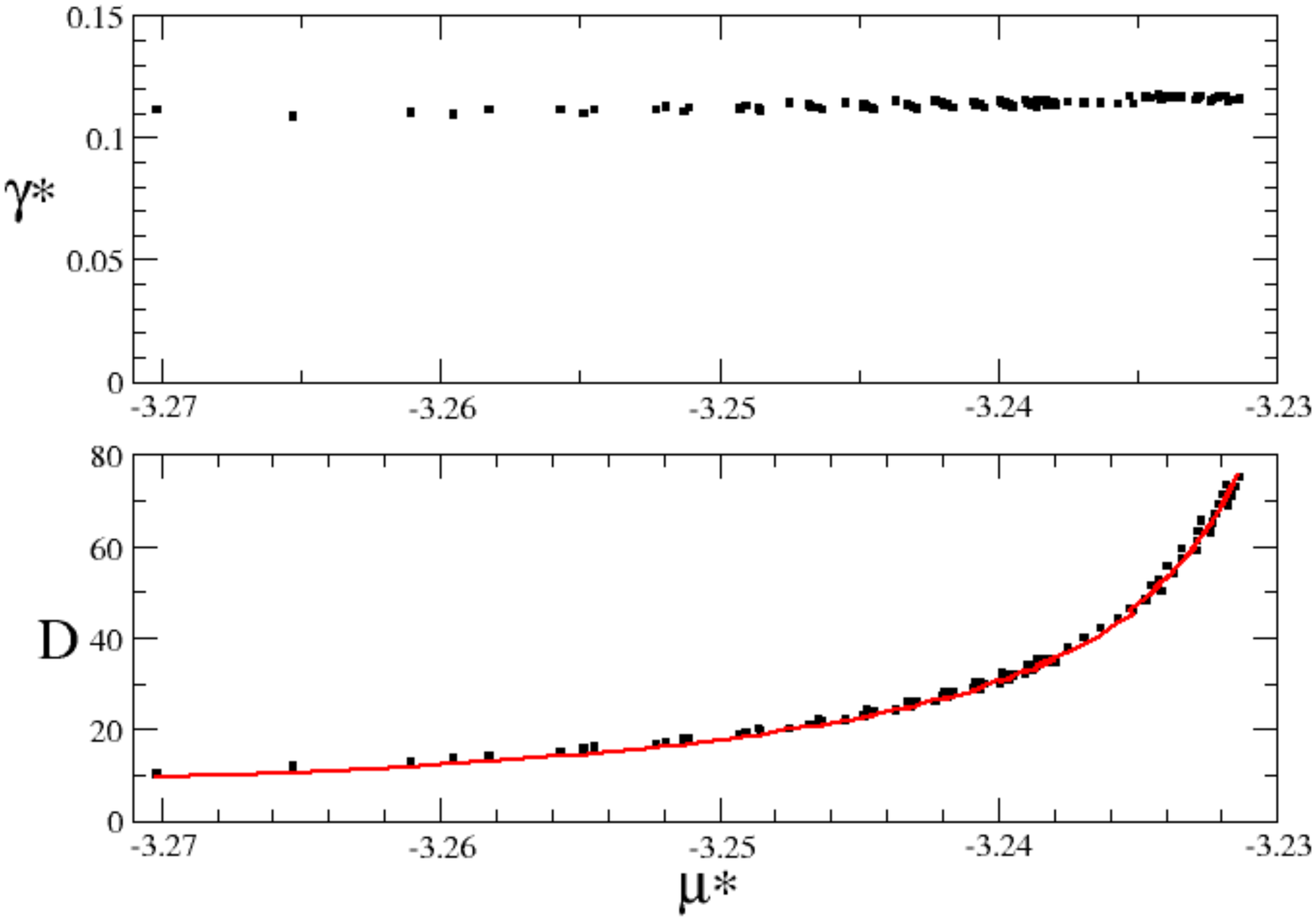}}
\caption{(color on line) Top: Surface tension $\gamma=\frac{L \ell f_{ex}}{\pi D}$ as a function of the chemical potential on the bubble branch. 
Bottom: Nucleus diameter $D$ versus chemical potential. The points are extracted from the density profiles (with the Gibbs dividing interface) 
and the red line is the prediction from the classical nucleation theory (see text).}
\label{fig_LMFT6} 
\end{figure*}

The same approach can be followed here, even if the system is confined in a pore, since one can easily estimate interfacial excess free energies. 
Consider the rail states, which correspond to the liquid/vapor coexistence in the pore at $\mu_{coex}$. If one defines the volume fraction $x$ 
of the liquid-like phase from $\bar \rho=x\rho_L(\mu_{coex})+(1-x)\rho_V(\mu_{coex})$, where the subscripts $L$ and $V$ denote the vapor 
and liquid phase respectively, the free energy density $\bar f=\bar F/V$ of the system 
can then be written as $\bar f=x f_L(\mu_{coex})+(1-x)f_V(\mu_{coex})+f_{ex}$ where $f_{ex}$ is the excess energy related to the presence 
of an interface between the two phases. We have checked that the quantity $\ell f_{ex}=2\gamma$ (the factor 2 comes from the presence of 
two liquid-vapor interfaces) is constant along all the rail states and equal to $0.23 w_{ff}$ (discarding the small oscillations due to 
lattice effects), see Fig.~\ref{fig_LMFT6}. The quantity $\gamma \simeq 0.117 w_{ff}$ has therefore 
the meaning of a surface tension in the $(y,z)$ plane between the vapor-like and liquid-like phases.  

How to use this value for determining nucleation barriers? The crucial point is the shape of the critical nucleus. Tallanquer 
et al.\cite{talanquer94} showed by local density-functional theory in the gradient approximation that the shape of the critical 
nucleus can either be attached to one of the walls or can bridge the two
walls, depending on the conditions. For a small separation between the walls, the 
critical nucleus bridges the walls and one can anticipate that nuclei have a round shape in the $(x,y)$ plane.  
Applying the classical nucleation theory with this hypothesis readily yields the value of the critical diameter 
$D^*$ of the droplet (resp. bubble) as a function of the difference $\Delta \omega$ between the 
grand-potential densities of the vapor-like (resp. liquid-like) GC metastable state and the liquid-like 
(resp. vapor-like) GC stable state at a given chemical potential $\mu$ : $D^*=\frac{2\gamma}{\vert \Delta \omega \vert}$. 
The corresponding energy barrier is $\Delta \Omega^*=\frac{\pi H \gamma^2}{\vert \Delta \omega \vert}$. These 
quantities may be rewritten close to the coexistence, where $\rho_L(\mu)-\rho_V(\mu)\simeq \rho_L(\mu_{coex})-\rho_V(\mu_{coex}) =\Delta 
\rho$, as $D^*=\frac{2\gamma}{\Delta \rho\vert \mu_{coex}-\mu \vert}$ and 
$\Delta \Omega^*=\frac{\pi H \gamma^2}{\Delta \rho \vert \mu_{coex}-\mu \vert}$.

\paragraph{The bubble and droplet branches.}
What is the relation between the critical nuclei theoretically described above and the observed bubbles and droplets? 
As can be seen in Fig.~\ref{fig_LMFT5} (and expected), the branches associated with the gas-like, liquid-like and rail states 
are independent of the system size, except for their 
extension in the $\rho-\mu$ diagram. On the other hand, those associated with the bubble and droplet states exhibit a 
simple system size dependence as explained now.

Consider for instance a state $(\mu, \bar \rho)$ on the bubble branch in a box $L\times \ell \times H$. Outside the 
bubble and its interface, the density is equal to the density of the liquid-like 
branch $\rho_L(\mu)$. If the bubble is sufficiently large, the density inside the bubble is equal to 
the density of the vapor-like branch $\rho_V(\mu)$. Taking advantage of the 
approximately  cylindrical symmetry  
of the bubble around the $z$-axis, one defines the diameter $D=D(\mu)=2R$ of 
the bubble as the diameter of the cylindrical surface dividing the two 
liquid-like and vapor-like phases, which ensures $\bar 
\rho(\mu)=x\rho_L(\mu)+(1-x)\rho_V(\mu)$ where $x=\frac{V_L}{V}$ and
$V_L$ the volume of the liquid-like phase. The free-energy density $\bar f$ of the 
system can then be written as $\bar f=x f_L(\mu)+(1-x)f_V(\mu)+f_{ex}$.  For large bubbles ($D>>1$), we have checked that 
$x \simeq \frac{\pi D^2}{4L\ell}$ and that $\frac{L \ell f_{ex}}{\pi D}$ is almost constant with the chemical potential (see Fig.~\ref{fig_LMFT6}) and 
equal to the surface tension $\gamma$ previously determined. (We 
have not found any noticeable influence of the interface curvature.)

Consistent with this approximation is that 
the distance in chemical potential from the liquid-vapor coexistence is small. The Kelvin equation should then be valid so that 
$\vert \mu-\mu_{coex}\vert =\frac{2\gamma}{\Delta \rho D}$ as  illustrated in Fig.~\ref{fig_LMFT6}: this means that the bubble (and droplet) 
branches are the locus of the critical nuclei introduced above.   
Finally, one finds  the expression of the bubble-branch equation,
\begin{equation}
\bar \rho (\mu)=\rho_L(\mu)-\frac{\pi}{HL\ell} \frac{H\gamma^2 }{\Delta 
\rho\vert \mu_{coex}-\mu \vert^2}\,,
\end{equation}
which makes their size dependence explicit. The curves all collapse when one plots $H L\ell \bar \rho(\mu)=N(\mu)$ 
as a function of $\mu$.

\paragraph{The morphological transitions.}

One cannot find analytically the density of the transition between the vapor-like 
and the droplet branch. Physically, the transition takes place when the adsorbed film can supply enough fluid to form the droplet.  
Data inspection shows that the smallest possible (stable) droplet is essentially 
independent of system size, but what about the equilibrium droplet at the 
transition? We have found that the size of the equilibrium droplet grows with the system size and, accordingly, that the 
corresponding chemical potential decreases to $\mu_{coex}$.

On the other hand, the transition between the droplet and rail states is 
essentially controlled by the geometry of the system. 
The excess free energies of these states are indeed the product of the surface tension $\gamma$ 
by the perimeter of the interface: $2\ell$ for the rail ($\ell<L$), $\pi D$ for the 
droplet. Equality of these excess free energies leads to the following values on the droplet branch at the transition: 
$(\bar \rho-\rho_V)/(\rho_L-\rho_V) = \frac{1}{\pi} \frac{\ell}{L}$ and  $\mu=\mu_{coex}+\frac{\pi \gamma}{H \Delta \rho \ell}$. 
Note that for square systems, $\rho$ at the transition does not depend on the system size, as illustrated in Fig.~\ref{fig_LMFT5}.
The same relation applies for the transition between the rail and the bubble 
states (with the exchange $L\leftrightarrow V$). These transitions can never be avoided: the system, for instance on adsorption, will 
always visit first the droplet state, then the rail state and finally the bubble state.

\subsection{Monte Carlo Results}

\subsubsection{Adsorption/desorption isotherms}

\begin{figure*}
\resizebox{1.7\columnwidth}{!} {\includegraphics{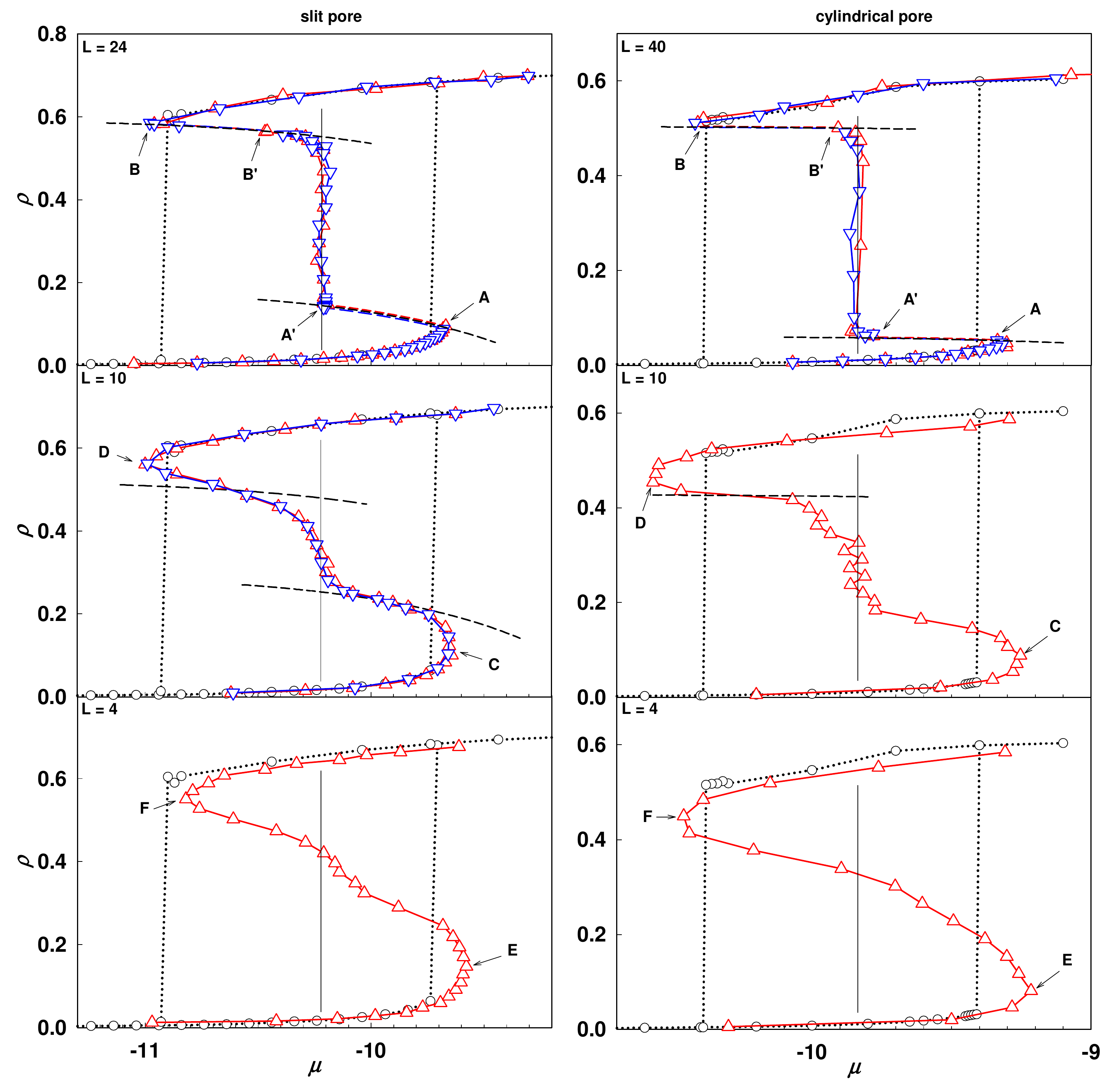}}
\caption{Left panel: Adsorption (up triangles) and desorption (down triangles) 
isotherms of a Lennard-Jones fluid in a slit pore of height $H = 6$, width $\ell 
= 8$, and lengths $L = 4, 10,$ and $24$ from bottom to top. Right 
panel: Same as left for a cylindrical pore of diameter $D = 6$ and lengths $L = 
4, 10,$ and $40$ from bottom to top. For the smallest lengths the desorption 
isotherms are indistinguishable from the adsorption and are therefore not 
shown. The relative size of the reservoir with respect to the pore is 66.7 for the slit pore and 50 for the 
cylinder; for these values, the conditions are close to canonical for the fluid 
in the pore. Labels A to F are used in subsequent figures. The open circles and dotted lines denote the grand-canonical 
adsorption and desorption isotherms obtained for the largest $L$. The dashed 
lines represent the canonical constraint for the total system comprising the 
pore and the reservoir for some specific points along the isotherms. The vertical 
thin lines materialize the coexistence points.}
\label{fig_MC1} 
\end{figure*}

The measured amount of fluid in the pore upon isothermal adsorption or 
desorption, as obtained in the atomistic model, is shown in Fig.~\ref{fig_MC1}. 
In both pore geometries (slit and cylinder) the pore length drastically 
influences the results. The pore length $L$ varies between 4, a value slightly 
smaller than its width or diameter, to significantly larger values ensuring high 
aspect ratios. The drastic changes occur for the smallest values of $L$. All 
isotherms then exhibit reversibility and a typical continuous $S$-shape. Increasing $L$ 
progressively leads to a deformation 
of the isotherms with the appearance of a steeper portion in the center. For large 
enough $L$, the isotherms exhibit two discontinuities (associated with 
morphological transitions) and three separated branches: the lowest-density one 
corresponds to a monolayer of fluid adsorbed at the walls; the highest-density one corresponds to a 
dense liquid-like filling the pore. In both cases the density profile describing the fluid is homogeneous in the 
direction parallel to $L$. On the other hand, the third branch of intermediate density corresponds to an 
inhomogeneous fluid state, made of fluid menisci or bridges anchored at the 
walls. 

Grand canonical simulations were performed in the largest systems, close to the 
thermodynamic limit (shown as circles and dotted lines in Fig.~\ref{fig_MC1}). 
As can be seen, the low- and high-density branches closely follow those obtained 
in the grand-canonical ensemble, except for the smallest system due to finite-size effects. 
Note also the small differences in the limits of stability, even in 
the largest systems. The mesoscopic canonical ensemble attenuates fluctuations, thereby 
stabilizing the metastable branches. 

What happens around the discontinuities? The total system comprising the pore 
and the reservoir is in the canonical ensemble. The corresponding constraint 
between the chemical potential and the amount of fluid in the pore appears as 
dashed lines. Note that if the pore were in the canonical ensemble, the 
constraint would result in horizontal dashed lines. In our case, the ratio 
between the reservoir size and the pore size (66.7 for the slit pore and 50 for 
the cylinder) is such that the pore is close to the canonical situation. As can 
be seen, for the largest systems, after the limit of stability of the low- and high-density branches are 
reached, the isotherm follows these constraints. 

\subsubsection{Grand potential calculations}

\begin{figure*}
\resizebox{1.7\columnwidth}{!} {\includegraphics{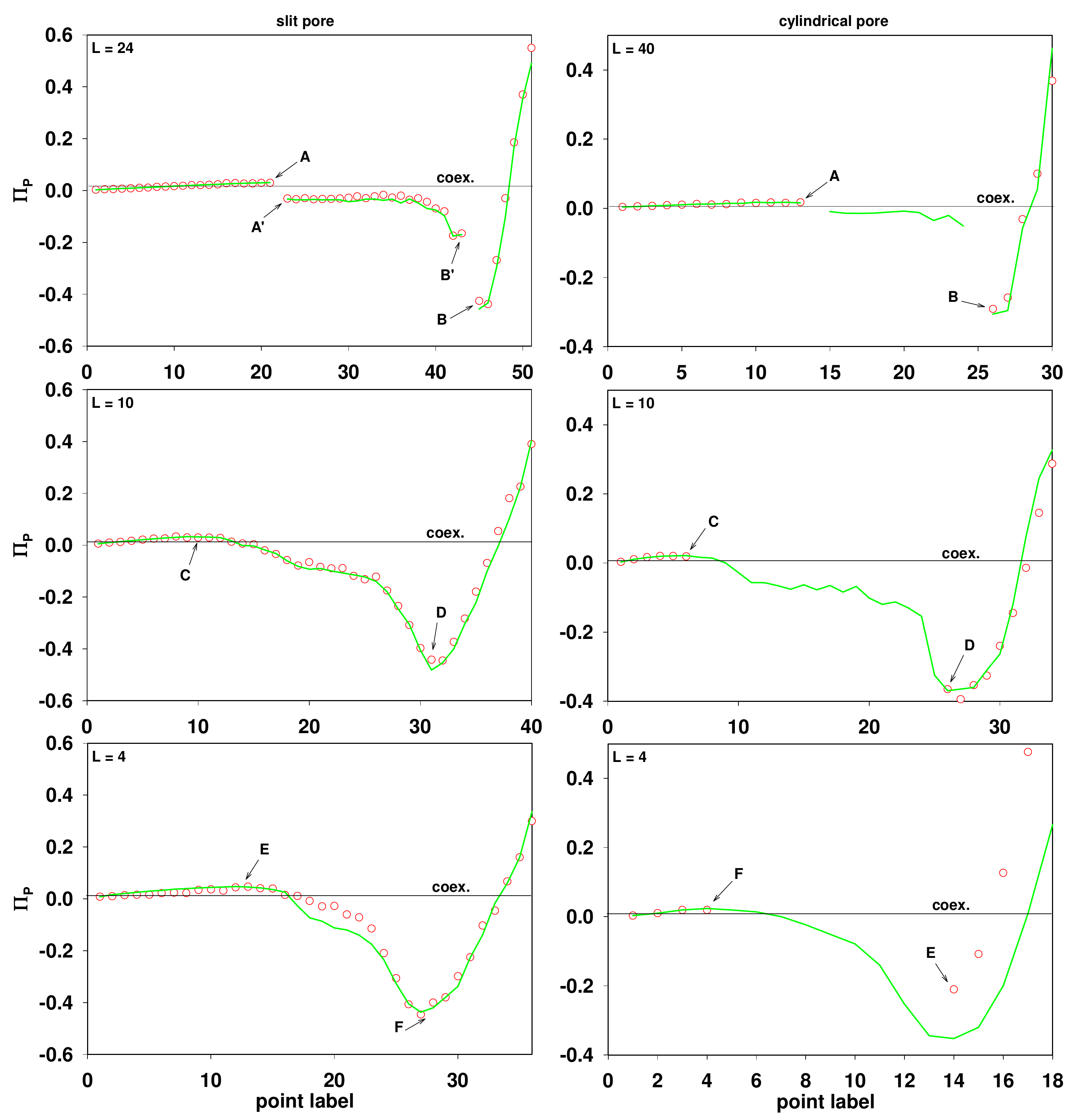}}
\caption{Grand-potential density, $\Pi_{\rm P} = - \Omega_{\rm P}/ V_{\rm P}$, of the fluid in the pore along 
the adsorption isotherm. The data are given as a function of the occurrence number of the points along the isotherm,
the points being labeled from the beginning of the adsorption 
in the gas regime to the full filling of the pore by the liquid. Labels A to F are the same as in
Fig~\ref{fig_MC1}. Solid lines: thermodynamic integration of the Gibbs relation $d\Pi = \rho d\mu$. Symbols: 
direct computation using Eq.~\ref{Eqn_direct} when possible (density profiles homogeneous in at least one direction, see text). 
Left panel: slit pore of height $H = 6$, width $\ell = 8$, and lengths $L = 4, 10,$ and $24$ from bottom to top. 
Right panel: cylindrical pore of diameter $D = 6$ and lengths $L = 4, 10,$ and $40$ from bottom to top. 
The horizontal thin lines materialize the coexistence value of the grand potential density in the thermodynamic limit.}
\label{fig_MC2} 
\end{figure*}

The fluid grand-potential density in the pore $\Pi_{\rm P}$ was calculated 
during the course of the simulation, by using the direct algorithm when possible 
[Eq.~(\ref{Eqn_direct})]. It was also computed by thermodynamic integration of 
the Gibbs relation $d\Pi = \rho d\mu$ 
along the adsorption isotherm. For clarity, the results are given as a function 
of the point label (points are labeled from the beginning of the adsorption in 
the gas regime to the full filling of the pore by the liquid) in order to avoid 
superimposition of points (see Fig.~\ref{fig_MC2}). Note that the applicability 
of the thermodynamic
integration is not straightforward since the adsorption/desorption isotherms are 
discontinuous (while reversible). By convention, discontinuities are integrated 
as single (large) increments. The associated error will be rigorously evaluated 
in next section. 

Let us first focus on the gas-like and liquid-like branches of the isotherms, which correspond to fluid 
configurations that are uniform along the pore axis. The thermodynamic results can be compared 
to the direct algorithm [Eq.~(\ref{Eqn_direct})]. A disagreement is observed between the two 
methods for the smallest cylindrical pore ($L = 4$), most probably due to trivial finite size effects.
Otherwise, for all other systems, the agreement is excellent, validating the thermodynamic integration
of the grand potential along the isotherms. This result is quite remarkable since the isotherms 
exhibit discontinuities. 

For the intermediate region of the isotherms, corresponding to a backward evolution of the chemical 
potential with the fluid density in the pore, the fluid configurations are 
inhomogeneous along the direction $L$, with a rail (resp. droplet) of liquid on 
the low-density side, and a cylinder (resp. bubble) of gas on the high-density 
side for the slit (resp. cylindrical) pore. As a consequence, the direct method 
is expected to fail and the corresponding results are not shown. However, the 
case of the slit pore can be considered more carefully. In this geometry, for a 
large enough length $L > \ell = 8$, visual inspection of the configurations shows that 
the orientation of the liquid rail or of the gas cylinder 
is always parallel to the direction $\ell$, \textit{ie}, the fluid is uniform along this direction. 
The direct method can thus be applied with the virtual variations performed along $\ell$. (We checked 
that for the homogeneous gas-like and liquid-like branches the results are identical to those 
obtained with virtual variations along $L$.) The interesting results are 
those obtained in the inhomogeneous intermediate region. As seen from 
Fig.~\ref{fig_MC2}, the thermodynamic 
integration results coincide with those obtained directly during the course of the 
simulation for $L = 10$ and $24$. For the smallest length $L = 4$ a disagreement appears. 
Visual inspection of the atomic configurations shows that the direction of the liquid rail 
or of the gas cylinder fluctuates between the $\ell$ and $L$ directions: as a consequence, the 
direct method actually fails, and the results should not be compared to thermodynamic integration. 

The global picture that arises is that for large enough systems (to avoid spurious finite-size effects), 
the thermodynamic integration is a rational route to obtain the grand potential. The analytical 
calculation of the errors associated with the discontinuities (see next section) 
validates this observation.  

\begin{figure*}
\resizebox{1.7\columnwidth}{!} {\includegraphics{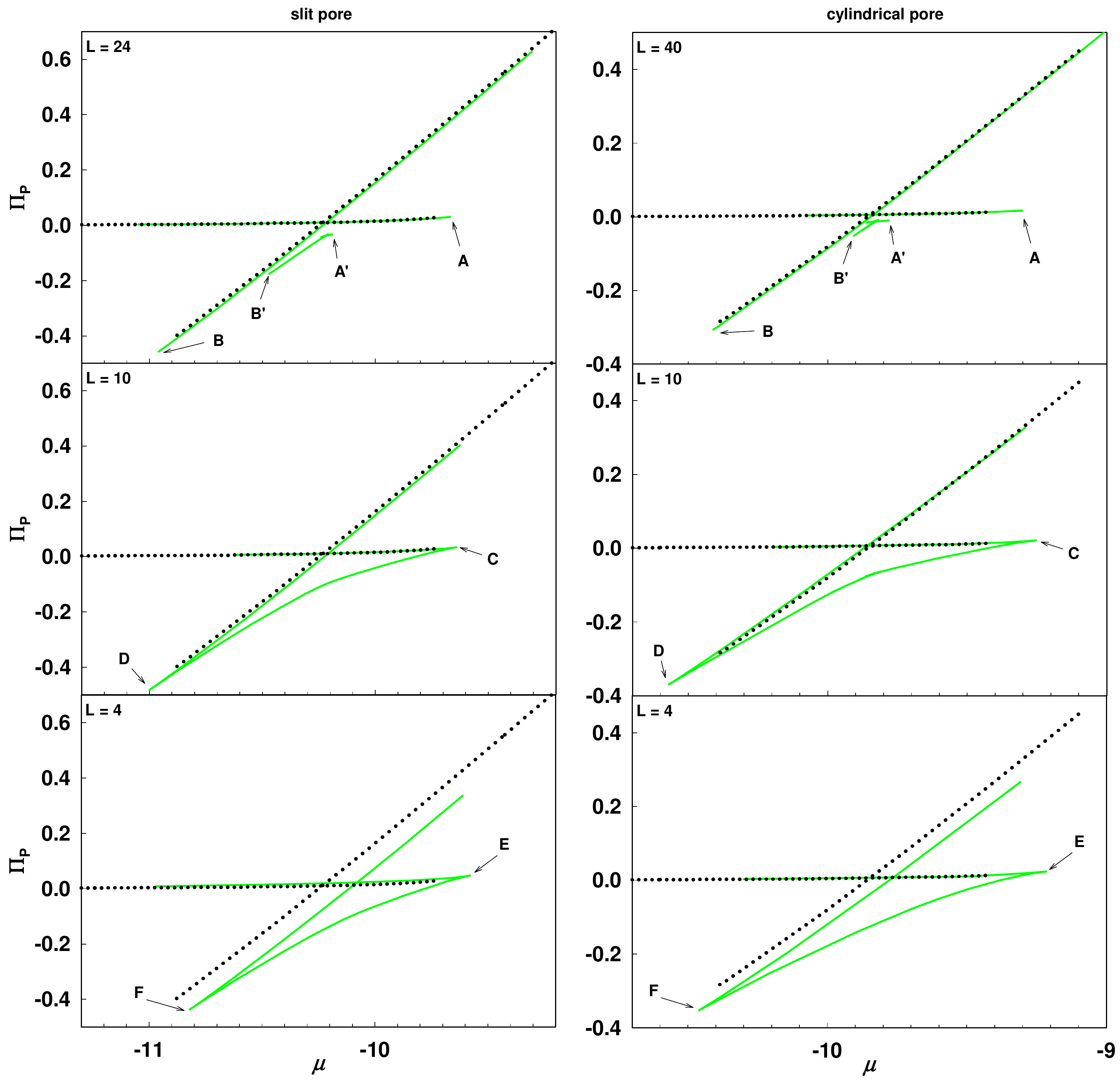}}
\caption{Grand-potential density, $\Pi_{\rm P} = - \Omega_{\rm P}/ V_{\rm P}$, of the fluid in the pore versus 
fluid chemical potential $\mu$. Labels A to F are the same as in
Fig~\ref{fig_MC1}. Solid lines: thermodynamic integration of the Gibbs relation $d\Pi = \rho d\mu$. Dotted lines: 
grand-canonical computation performed in a pore long enough  
for representing the thermodynamic limit. 
Left panel: slit pore of height $H = 6$, width $\ell = 8$, and lengths $L = 4, 10,$ and $24$ from bottom to top. 
Right panel: cylindrical pore of diameter $D = 6$ and lengths $L = 4, 10,$ and $40$ from bottom to top.
}
\label{fig_MC3} 
\end{figure*}

Fig.~\ref{fig_MC3} (solid lines) shows the evolution of the grand-potential 
density obtained by thermodynamic integration versus the chemical potential $\mu$. 
There are two principal branches with very different slopes (given 
by the density of the adsorbed fluid), corresponding to the gas-like and 
liquid-like portions of the isotherms. These branches cross at coexistence 
between the gas-like and liquid-like states. For the lowest values of $L$, these 
branches are connected by a continuous line corresponding to the backward 
portion of the isotherm. On the other hand, for the largest values of $L$, the line 
associated with the inhomogeneous intermediate states appear as a separate branch 
just below the coexistence point. The difference in grand potential between 
these states and the coexistence point is due to the contributions of the two 
gas/liquid interfaces of the nonuniform states (the corresponding grand 
potential is higher). The numerical calculations give a reduced surface 
tension of 1.25 for both systems and for $L \geq  10$.

The grand-potential density obtained by thermodynamic integration can be 
compared to a grand-canonical computation performed on the largest pores 
(that are closer to the thermodynamic limit). The results of the grand-canonical 
computation are available only for the gas-like and liquid-like branches (see 
Figs.~\ref{fig_MC1} and \ref{fig_MC3}, dotted lines). Fig.~\ref{fig_MC3} shows that, along these 
branches, the agreement with the results of thermodynamic integration is 
excellent for the largest length and for the two geometries. The agreement 
however deteriorates as one considers smaller lengths for the thermodynamic 
integration. It is not a failure of the thermodynamic integration, but finite-size effects then affect the calculations. 
This is clearly visible in the isotherms (see Fig.~\ref{fig_MC1}).

\subsubsection{Thermodynamic integration}

The question we now address follows from the previously observed numerical agreement 
between the grand-potential calculations and the thermodynamic integration 
of the adsorption isotherms. Is it valid to calculate the fluid ``grand 
potential'' in the pore $\Omega_{\rm P}$ by a direct integration of 
$\left<N_{\rm P}\right> (\mu)$ along the isotherm when the latter is reversible yet 
discontinuous? Note that the relation we consider here involves the average number of atoms 
in the pore, not the total number of atoms in the whole system. Integration is 
obviously possible along the continuous portions of the adsorption branches. A 
difficulty arises when discontinuities (gaps) are present, which may then 
introduce sizeable errors and even invalidate the procedure. By using the fact that 
the total number of particles $N$ and the total free energy $F$ are the same on 
both sides (labeled 1 and 2) of a given gap, one has from Eq.~\ref{Eqn_Omg}:
\begin{eqnarray}
{\Omega_{P,1} - \Omega_{P,2}} && = -( \mu_{2} - \mu_{1}) N - \left( \Omega_{R,2} 
- \Omega_{R,1} \right) \nonumber \\
&& = -(\mu_{2} - \mu_{1}) \frac{\left< N_{P,2} \right> + \left<N_{P,1}\right>} 
{2} - \Delta
\label{Eqn_Int1}
\end{eqnarray}
with
\begin{eqnarray}
\Delta = (\mu_{2} - \mu_{1}) \frac{\left<N_{R,2}\right> +  \left<N_{R,1}\right>} 
{2} + \left( \Omega_{R,2} - \Omega_{R,1} \right) {\rm .}
\label{Eqn_Int2}
\end{eqnarray}
In these expressions, $P$ and $R$ denote the pore and reservoir contributions. 

The term $-(\mu_{2} - \mu_{1}) \frac{\left< N_{P,2} \right> + \left<N_{P,1}\right>} {2}$ of Eq.~\ref{Eqn_Int1} 
exactly corresponds to what would be 
obtained by applying the usual thermodynamic integration of $\left<N_{\rm 
P}\right> (\mu)$ via a discrete integration performed over simulation points, in 
which the discontinuity $1-2$ is considered as a single increment. $\Delta$ is 
the difference between the grand potential calculated by thermodynamic 
integration and the true one. To evaluate this term we use the ideal-gas expressions 
for $N_{\rm R}(\mu)$ and $\Omega_{\rm R} (\mu)$. An expansion in terms of the 
chemical potential difference $\mu_{2} - \mu_{1}$ then leads to
\begin{eqnarray}
\Delta = (\mu_{2} - \mu_{1})^{3} \frac{\left<N_{R,1}\right>} {12 k_B^{2} T^{2}} + 
O((\mu_{2} - \mu_{1})^{5})\,,
\label{Eqn_Int3}
\end{eqnarray}
or alternatively in terms of the difference in adsorbed amount 
$\left<N_{P,2}\right> - \left<N_{P,1}\right> = \left<N_{R,1}\right> - 
\left<N_{R,2}\right>$, to
\begin{eqnarray}
\Delta = (\left<N_{P,1}\right> - \left<N_{P,2}\right>)^{3} \frac{k_B T} {12 
\left<N_{R,1}\right>^{2}} \nonumber \\ + O(( \left< N_{P,1} \right> - \left< 
N_{P,2} \right> )^{5}) {\rm .}
\label{Eqn_Int4}
\end{eqnarray}
Note that either in the canonical ensemble ($\left< N_{P,1} \right> = \left< 
N_{P,2} \right> $) or in the grand-canonical ensemble ($\mu_{1} = \mu_{2}$), the 
integration is exact ($\Delta = 0$ to all orders). In a generic mixed (or mesoscopic-canonical) ensemble, 
the integration is not exact, but the above equations show that the standard  
thermodynamic integration of $\left<N_{\rm P}\right> (\mu)$ gives the grand 
potential within an error of order 3 in the gap $1-2$. If the observed gaps 
are small, and in particular not much larger than the increments between two 
simulation points, the error introduced is negligible compared to the second 
order error introduced by the discrete integration procedure. 

The last point to be discussed is what happens if the fluid in the reservoir is 
not ideal, as for instance close to the critical point? In this case, the last 
term $\left( \Omega_{R,2} - \Omega_{R,1} \right)$ in Eq.~(\ref{Eqn_Int1}) should 
be replaced by $\int_{1}^{2}d \mu \left<N_{\rm R}\right> (\mu)$, which corresponds  
to a thermodynamic integration of the fluid properties from state 1 to state 2 in 
the whole system. This shows that the error introduced by the thermodynamic 
integration of $\left<N_{\rm P}\right> (\mu)$ is the difference between the 
continuous and the discrete integral of $\left<N_{\rm R}\right> (\mu)$ between 
points 1 and 2 and therefore is now of order 2 in the gap $1-2$. (Note that if 
this gap is of the order of the increment between simulation points, the error 
is of the same order of magnitude as the one due to the discrete integration.)

We can summarize the above discussion as follows: If the isotherm obtained in 
the mixed (mesoscopic-canonical) ensemble is reversible and exhibits small gaps, thermodynamic 
integration can be safely performed to calculate the grand potential of the 
confined fluid. The size of the pore must be large enough for ensuring that the 
result of thermodynamic integration in a finite-size system coincides with the 
thermodynamic limit. If the gaps are large but the fluid is ideal in the 
reservoir (far from the critical point for instance), the error is expected to 
be of order 3 in $\mu_{2} - \mu_{1}$ or $\left< N_{P,1} \right> - \left< N_{P,2} 
\right>$, and thus quite small. If the fluid is not ideal in the reservoir, the 
error is of second order. In the latter case, determining the total Helmholtz free energy $F$ by 
thermodynamic integration and using Eq.~(\ref{Eqn_Omg}) is then a better route.

\section{Conclusion}

In this paper we have studied the influence of the system size on the properties of a fluid adsorbed in a nanopore, when the 
reservoir is small enough to enforce canonical or near-canonical ensemble conditions for the adsorbed fluid. The system here 
refers to the pore, which is chosen of simple, slit or cylinder, geometry. We have done so by 
means of two complementary methods: the density-functional theory for a lattice-gas model and Monte Carlo simulations of an atomistic 
model. The motivation behind the study was two-fold: first, a theoretical investigation of the inhomogeneous fluid states that 
can be stabilized in the canonical ensemble for a finite-size system and, second, an assessment of the ``gauge-cell method''\cite{neimark02} 
that was devised for computing thermodynamic quantities in numerical studies.

Within the density-functional approach, we have obtained results that can be summarized as follows:

\begin{itemize}
\item The path followed by the system on adsorption and desorption depends crucially on the system size and
 on the degrees of freedom  allowed for the density profiles.
\item Large systems exhibit morphological transitions between ``homogeneous'' phases (more precisely, phases having the 
symmetry imposed by the geometry of the pore) and inhomogeneous phases (where the symmetry is broken). Hysteresis can 
moreover be suppressed and equilibrium found since the density-functional theory provides the free-energy values.
\item Among the branches of the isotherms associated with inhomogeneous fluid phases, the droplet and bubble branches are 
found as the loci of the critical nuclei with properties that are consistent with the classical nucleation theory.
\item Continuous and reversible isotherms are observed in small systems. However, when increasing system size, the symmetries of 
the external potential imposed by the pore can be broken for the density profiles and we have devised a technical trick to obtain 
such profiles in a density-functional theory.
\end{itemize}

On the other hand, via the Monte Carlo simulations: 

\begin{itemize}
\item We have focused on the idea, which is behind the gauge-cell method, of using small systems 
to get a continuous van der Waals-like isotherm from which thermodynamic 
integration could be safely implemented. We have found that when going to the small system sizes required to 
observe continuous isotherms, the  thermodynamic properties are strongly affected by the limited size, especially in computer 
simulations where the required system size becomes smaller than the range of the interactions. Indeed, 
inhomogeneous states persist down to quite small system sizes so that the 
recovery of an $S$-shape quasi-continuous isotherm is very slow when decreasing the size. This procedure is therefore 
not a good way to extract the thermodynamic properties of the adsorbed fluid.
\item We have suggested an alternative method, in which the system (pore) size is reduced only until a 
reversible isotherm is found, even if discontinuities are present. The size is larger than in the above case and the properties of the 
gas-like and liquid-like branches of the isotherms are not altered with respect to the thermodynamic limit. We have shown 
that thermodynamic integration is still valid in the presence of jumps, provided that the isotherm is reversible. This is an extension 
of the gauge-cell method to cases where inhomogeneous states and morphological transitions are present.
\end{itemize}

We finally note that our study supports the view that playing with system and reservoir size (as well as with temporary modifications 
of the external potential exerted by the pore on the fluid) seems to be a very generic route to numerically investigate 
nucleation and the associated free-energy barriers in adsorbed fluids.

\end{document}